# Propagation of a Bessel beam in conducting media

D. Mugnai
"Nello Carrara" Institute for Applied Physics, CNR Florence Research Area,
Via Madonna del Piano 10,
50019 Sesto Fiorentino, Italy

The aim of this note is to analyze the propagation of a Bessel beam coming from a dielectric medium (say air), and impinging into a conducting medium, se. We will demonstrate that the beam does not change its shape when propagating in the conducting medium, but solely attenuates going father from the interface.

Let us start considering a system formed by two half-space, 1 and 2, as sketched in Fig. 1.

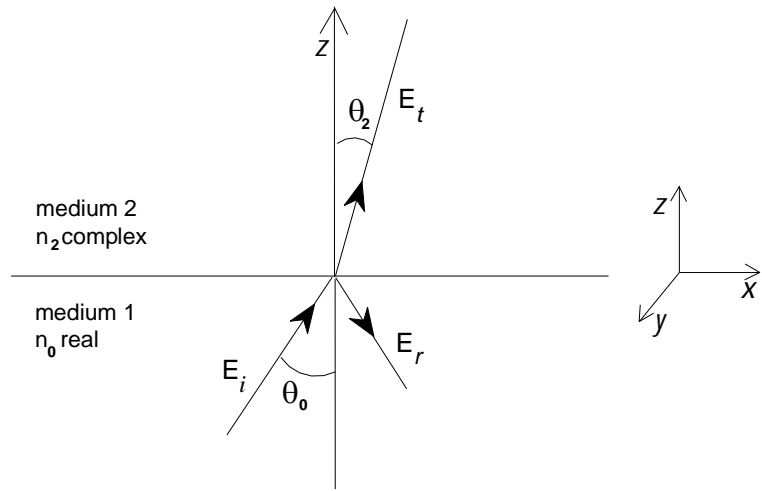

Fig 1. Scheme of propagation: the Bessel beam $E_i$ coming from medium 1, which is a perfect dielectric, impinges in medium 2, which is a conducting medium. The refractive index $n_0$ is real, while $n_2$ is complex; $\theta_0$ and $\theta_2$ are the incident (real) and refractive (complex) angles, respectively; $E_r$ and $E_t$ refer to the reflected and transmitted fields, respectively.

A plane wave coming from medium 1 impinges on medium 2 with the incident angle $\theta_0$. Be medium 1 a perfect dielectric and medium 2 a conducting medium, separated from medium 1 by a plane interface. The plane wave in medium 1 is

[1] $$E_i = \exp[ik_0(\alpha x + \beta y + \gamma z)]$$

This impinging plane wave is expected to give rise in medium 2 to a plane wave, but of dissociate (or evanescent, or complex) type, due to the fact that $n_2$ is complex, hence an exponential decrease of the real amplitude is expected. Thus in medium 2 we write

[2] $$E_t = T\exp[ik_2(\alpha_2 x + \beta_2 y + \gamma_2 z)] = T\exp[ik_0 n_2(\alpha_2 x + \beta_2 y + \gamma_2 z)], \quad (n_2 \text{ complex number})$$

where $k_0$ and $k_2 = (n_2 k_0)$, are the wave numbers in medium 1 (vacuum or air) and in medium 2, respectively; $T$ is the transmission coefficient; $\alpha$, $\beta$, $\gamma$ are the real cosine directors of the direction of propagation in medium 1, while $\alpha_2, \beta_2$, and $\gamma_2$ are constants only subject to the condition that

$$\alpha_2^2 + \beta_2^2 + \gamma_2^2 = 1.$$

For the sake of simplicity we have taken the amplitude of the incoming wave equal to one, and we have omitted the temporal factor $\exp(i\omega t)$.

In a system of spherical coordinates the quantities $\alpha$, $\beta$, $\gamma$ may be written as

$$\alpha = \sin\theta_0 \cos\psi, \beta = \sin\theta_0 \sin\psi, \gamma = \cos\theta_0,$$

and

[3] $$\alpha_2 = \sin\theta_2 \cos\psi_2, \beta_2 = \sin\varsigma_2 \sin\psi_2, \gamma_2 = \cos\theta_2,$$

where both $\theta_2$ and $\psi_2$ could in principle be complex angles.

However, it is not so, as can be inferred by simple physical considerations. As known [1], in the propagation through a plane interface into a conducting medium the surfaces of constant amplitude must be parallel to the interface. Thus, in order to meet this condition the products $n_2\alpha_2$ and $n_2\beta_2$ in Eq. [2] have to be both real, otherwise the constant-amplitude surfaces would be plane, but with an inclination with respect to the interface

Since the products $n_2\alpha_2$ and $n_2\beta_2$ are real, even $\tan\psi_2$ is real, as ratio of two real numbers:

$$\tan\psi_2 = \frac{n_2\beta_2}{n_2\alpha_2} = \frac{n_2 \sin\theta_2 \sin\psi_2}{n_2 \sin\theta_2 \cos\psi_2}$$

Hence $\psi_2$ is a real angle.

Now, by equating on the interface $z = 0$ the phase distribution of the impinging wave to that of the refracted wave, and choosing $\psi_2 = \psi$, we find the Snell law, that is

[4] $$n_2 \sin\theta_2 = \sin\theta_0 \quad (n_0 = 1).$$

It turns out that $\theta_2$ is complex:

$$\sin\theta_2 = \frac{1}{n_2}$$

Coming back to Eq. [2], it is expedient to refer to a cylindrical coordinates system $(\rho, \vartheta, z)$ such that

$$x = \rho\cos\varphi, \quad y = \rho\sin\varphi, \quad z = z.$$

Thus, Eqs. [1] and [2] can be written as

[5]
$$E_i = \exp[ik_0\rho(\sin\theta_0\cos\psi\cos\varphi + \sin\theta_0\sin\psi\sin\varphi)]\exp(ik_0\cos\theta_0 z)$$
$$= \exp(ik_0\cos\theta_0 z)\exp[ik_0\rho(\sin\theta_0\cos(\varphi-\psi))],$$

and

[6]
$$E_t = T\exp[ik_0\rho(n_2\sin\theta_2\cos\psi\cos\varphi + n_2\sin\theta_2\sin\psi\sin\varphi)]\exp(ik_0 n_2\cos\theta_2 z)$$
$$= T\exp(ik_0 n_2\cos\theta_2 z)\exp[ik_0\rho(\sin\theta_0\cos(\varphi-\psi))],$$

where use has been made of Eq.[4].

The quantity $n_2\cos\theta_2$ controlling the z-dependence of the wave is complex:

$$n_2\cos\theta_2 = \sqrt{n_2^2 - \sin^2\theta_0},$$

which indicates that the wave attenuates in the direction normal to the interface (z-axis).

Let us now pass to consider a Bessel beam. As known, a Bessel beam (also known as Bessel X-wave) originates from the interference of an infinite number of plane waves whose directions of propagation make the same angle, say $\theta_0$, with a given axis, say z.

Let us suppose to have in medium 1 such plane waves. Thus, the field propagating in medium 1 is given by the sum of an infinite number of waves like that of Eq. [5] and results in

[7]
$$E_i = \exp(ik_0 z\cos\theta_0)\int_0^{2\pi}\exp[ik_0\rho(\sin\varsigma_0\cos(\varphi-\psi))]d\psi$$
$$= 2\pi\, J_0(k_0\rho\sin\theta_0)\exp(ik_0 z\cos\theta_0),$$

where $J_0(k_0\rho\sin\theta_0)$ denotes the zero-order Bessel function of first kind.

Similarly, by integrating Eq. [6] between 0 and $2\pi$, we are able to obtain the field propagating in medium 2:

[8]
$$E_t = T\exp(ik_0 n_2 z\cos\theta_2)\int_0^{2\pi}\exp[ik_0\rho(\sin\theta_0\cos(\varphi-\psi))]d\psi$$
$$= 2\pi T\, J_0(k_0\rho\sin\theta_0)\exp(ik_0 n_2 z\cos\theta_2).$$

Looking at Eqs. [7] and [8], we arrive at this notable result: a Bessel beam propagating from a perfect dielectric medium to a conducting one does not change its shape in the passage. The field in the conducting medium still has a Bessel beam shape, even if its amplitude suffers some modification: at $z = 0$, it complex amplitude is not 1 but T, then it attenuates going away from the interface, due to the complex exponential factor.

## References

[1] - J. A. Stratton, "*Electromagnetic Theory*", McGraw-Hill, New York, 1941, Sec. 9.8.